\documentclass[%
reprint,
superscriptaddress,
 amsmath,amssymb,
aps,
prb,
]{revtex4-1}

\usepackage{graphicx}% Include figure files
\usepackage{dcolumn}% Align table columns on decimal point
\usepackage{bm}% bold math
\usepackage[colorinlistoftodos]{todonotes}
\begin{document}
\title{Local electronic structure of rutile RuO$_2$}

\author{Connor A. Occhialini}
\email{caocchia@mit.edu}
\affiliation{Department of Physics, Massachusetts Institute of Technology, Cambridge, MA 02139, USA.}
\author{Valentina Bisogni}
\affiliation{National Synchrotron Light Source II, Brookhaven National Laboratory, Upton, NY 11973, USA.}
\author{Hoydoo You}
\affiliation{Materials Science Division, Argonne National Laboratory, Argonne, IL 60439, USA.}
\author{Andi Barbour}
\affiliation{National Synchrotron Light Source II, Brookhaven National Laboratory, Upton, NY 11973, USA.}
\author{Ignace Jarrige}
\affiliation{National Synchrotron Light Source II, Brookhaven National Laboratory, Upton, NY 11973, USA.}
\author{J. F. Mitchell}
\affiliation{Materials Science Division, Argonne National Laboratory, Argonne, IL 60439, USA.}
\author{Riccardo Comin}
\email{rcomin@mit.edu}
\affiliation{Department of Physics, Massachusetts Institute of Technology, Cambridge, MA 02139, USA.}
\author{Jonathan Pelliciari}
\email{pelliciari@bnl.gov}
\affiliation{National Synchrotron Light Source II, Brookhaven National Laboratory, Upton, NY 11973, USA.}

\date{\today}

\begin{abstract}
Recently, rutile RuO$_2$ has raised interest for its itinerant antiferromagnetism, crystal Hall effect, and strain-induced superconductivity. Understanding and manipulating these properties demands resolving the electronic structure and the relative roles of the rutile crystal field and $4d$ spin-orbit coupling (SOC). Here, we use O-K and Ru $M_3$ x-ray absorption (XAS) and Ru $M_3$ resonant inelastic x-ray scattering (RIXS) to disentangle the contributions of crystal field, SOC, and electronic correlations in RuO$_2$.  The locally orthorhombic site symmetry of the Ru ions introduces significant crystal field contributions beyond the approximate octahedral coordination yielding a crystal field energy scale of $\Delta(t_{2g})\approx 1$ eV breaking the degeneracy of the $t_{2g}$ orbitals. This splitting exceeds the Ru SOC ($\approx160$ meV) suggesting a more subtle role of SOC, primarily through the modification of itinerant (rather than local) $4d$ electronic states, ultimately highlighting the importance of the local symmetry in RuO$_2$.  Remarkably, our analysis can be extended to other members of the rutile family, thus advancing the comprehension of the interplay among crystal field symmetry, electron correlations, and SOC in transition metal compounds with the rutile structure. 
\end{abstract}

%The locally orthorhombic symmetry of the Ru ions splits the $4d$ orbital levels beyond the octahedral contribution, introducing a new energy scale ($\Delta(t_{2g})\approx1$ eV) acting on the $t_{2g}$ subspace.

\maketitle

%\tableofcontents

%%INTRODUCTION%%

\section{Introduction}

Transition metal dioxides of the rutile structure exhibit many paradigmatic electronic phenomena arising from the delicate balance of strong electron correlations and spin-orbit coupling (SOC).  Few examples include the metal-insulator transition in VO$_2$ \cite{Goodenough1971,Goodenough1971a,He2016,Hiroi}, the half-metallic ferromagnetism of CrO$_2$ \cite{Korotin1998}, and the SOC-mediated spin Hall effect in IrO$_2$ \cite{Fujiwara2013,Sun2017,Sinova2015}. Many of these phenomena stem from the unique rutile structural symmetry, wherein transition metal sites of orthorhombic ($D_{2h}$) site-symmetry are coordinated to distorted oxygen octahedra, forming a bonding network with mixed edge- and corner-sharing octahedral configurations. The importance of this reduced symmetry was recognized early on \cite{Goodenough1971,Goodenough1971a, Sorantin1992}; however, an analysis that treats the SOC, electronic correlations, and reduced symmetry crystal field on equal footing is, so far, absent. Thus, accounting for the mixed degree of itinerancy and localization among the active $d$-orbitals is essential for understanding the origins of their structural, transport and magnetic properties. 

In the context of rutile oxides, RuO$_2$ is a special case due to intermediate $4d$ SOC and electronic correlation strength.  RuO$_2$ has long been regarded as a Pauli paramagnetic semi-metal, with early studies focusing on band structure descriptions of transport properties and optical/photoemission spectra \cite{Mattheiss, Glassford1994, Lin2004, Krasovska1995, DeAlmeida2006, Cox1986}. Only recently was it realized that the Fermi surface of RuO$_2$ exhibits a propensity for an itinerant antiferromagnetic (AFM) ground state \cite{Berlijn2017, Ping2015}. Subsequently, the  antiferromagnetism has been confirmed and room temperature collinear AFM order has been observed \cite{Berlijn2017,Zhu2019}.  Finally, the rutile symmetry, magnetic order, and SOC in RuO$_2$ have elicited further interest on transport properties that led to the observation of spin \cite{Sun2017,Jovic2018} and crystal \cite{Smejkal2020,Feng2020} Hall conductivities, and recently, to the discovery of strain-induced superconductivity \cite{Ruf2020,Uchida2020}.

A microscopic understanding of these properties in the $4d$/$5d$ rutile systems demands knowledge on the precise role of SOC and crystal field splitting \cite{Sun2017,Jovic2018, Xu2019,Ping2015,Kahk2014,Hu2000,Hirata2013,Clancy2012,Kim2018}. On this front, resonant inelastic x-ray scattering (RIXS) is an ideal tool for investigating $d$-orbital levels including relative SOC and orbital energetics \cite{DeGroot2008,MorettiSala2011,Bisogni2016,Sala2014}. While significant progress has been made in the study of $4d$ oxides using $L$-edge RIXS (in the tender x-ray regime) \cite{Suzuki2019,Gretarsson2019}, $M_{2,3}$-edge ($3p\rightarrow 3d/4d$) RIXS (in the soft x-ray regime) is in many ways analogous to the former \cite{Miedema2019,Wray2015,Chiuzbaian2005} and offers an alternative for investigating $4d$ physics \cite{Lebert2020}.

Here, we study RuO$_2$ ($4d^4$) with O $K$-edge ($1s\rightarrow 2p$) and Ru $M_3$-edge ($3p\rightarrow 4d$) x-ray absorption spectroscopy (XAS), Ru $M_3$-edge RIXS, and multiplet calculations. Our O $K$-edge XAS quantifies the octahedral ($O_h$) crystal field component $\Delta(\underline{e_g} - \underline{t_{2g}})$ to $\approx2.6$ eV and identifies a pronounced polarization anisotropy in the oxygen ligands associated with the Ru-O bonding network.  Despite the delocalized nature of the $4d$ states, our RIXS measurements uncover clear Raman-like $dd$-excitations allowing us to resolve the local orbital levels.  Our combined experimental evidence and crystal field multiplet (CFM) simulations underscore the dominant role of lower-symmetry (below $O_h$) crystal field splitting  ($\Delta(t_{2g})\approx1$ eV) over the $4d$ SOC (160 meV), which has the effect of breaking the $t_{2g}$ degeneracy. This defines how the orbital and band degeneracies are lifted from a high-symmetry coordination due to structural rather than relativistic effects, marking important constraints to explain the unconventional properties in RuO$_2$. More broadly, the intermediate nature of RuO$_2$ with respect to crystal field, SOC, and electron filling permits a clear comparison of the local electronic structure across the $3d$/$4d$/$5d$ rutile systems, which reveals striking and unexpected similarity, verifying the universal role of low-symmetry effects in rutile oxides \cite{Kahk2014, He2016,Ping2015}.  

We structured our work in the following sections: in Section \ref{experiment} we discuss XAS and RIXS experimental data, in Section \ref{modeling} we introduce the different theoretical frameworks and how they compare with experimental data, in Section \ref{discussion} we discuss the implications of our theory and data, and in Section \ref{conclusion} we summarize our work.

\section{Experiment}\label{experiment}

%% FIGURE 1 %%

\begin{figure}[t]
\includegraphics[width=\columnwidth]{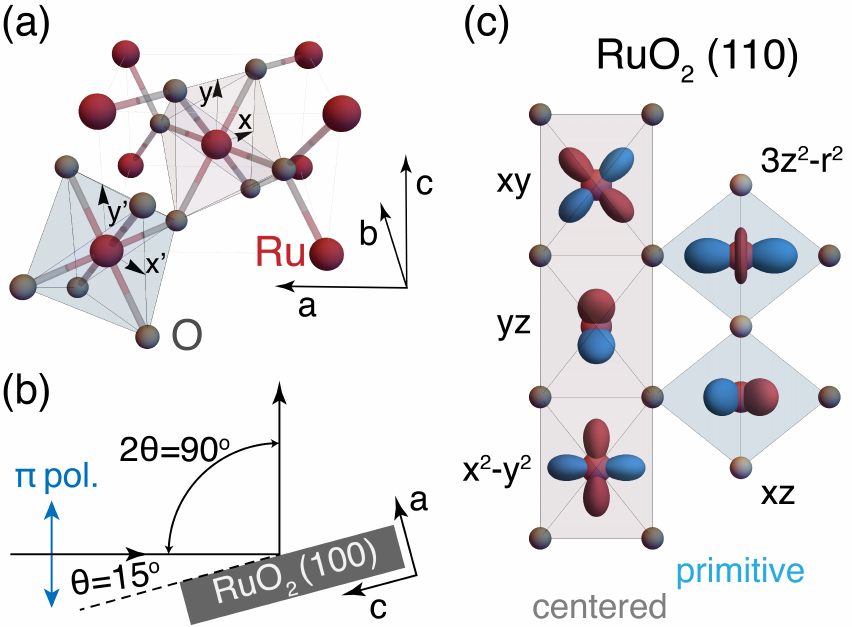}
\caption{(a) The rutile crystal structure of RuO$_2$, denoting the local Ru site axes, $xy$ and $x'y'$, used for orbital definitions at the two Ru lattice sites and (b) the scattering geometry used for all experiments. (c) A depiction of the $4d$ orbitals at the centered and primitive Ru lattice sites as viewed in the RuO$_2$ (110) plane, where the $xy$ and $3z^2-r^2$ orbitals belong to the $e_g$ set and the $xz$/$yz$/$x^2-y^2$ to the $t_{2g}$ set.}
\label{fig:fig1}
\end{figure}

A large, high-quality single crystal of RuO$_2$ with a well-oriented and polished $(100)$ facet was used for all measurements \cite{Zhu2019,Lister2002}.  The rutile crystal structure is shown in Fig. \ref{fig:fig1}(a) with the scattering geometry used for all measurements depicted in Fig. \ref{fig:fig1}(b). Figure \ref{fig:fig1}(a) defines both the crystallographic and local orbital axes at the two (centered and primitive) Ru sites, which are used to define the real $4d$ orbitals in Fig. \ref{fig:fig1}(c).  XAS linear dichroism at the oxygen $K$-edge and RIXS experiments at the ruthenium $M_3$-edge ($530$ eV and $463$ eV, respectively) were performed at the 2ID-SIX beamline at NSLS-II, Brookhaven National Laboratory (USA) \cite{Dvorak2016}.  XAS measurements were collected in Total Fluorescence Yield (TFY) with linear horizontal ($\pi$) and vertical ($\sigma$) polarizations and all measurements utilized a combined energy resolution of $\sim 125$ meV. 

%% FIGURE 2 %%

\begin{figure}[h]
\includegraphics[width=1.0\columnwidth]{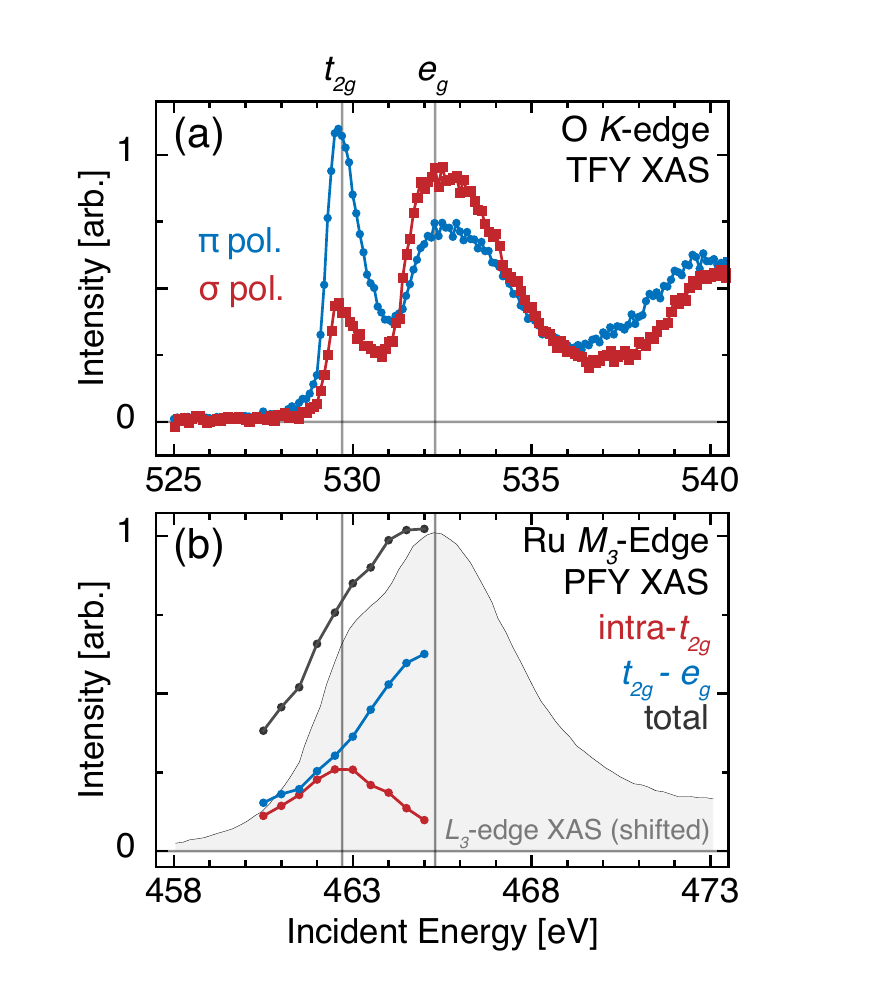}
\caption{(a) Oxygen $K$-edge XAS recorded in TFY with $\sigma$ (red) and $\pi$ (blue) incident polarization, as defined by the scattering plane in Fig. \ref{fig:fig1}(b).  (b) Ru $M_3$-edge XAS with $\pi$ incident polarization recorded in PFY (dark grey data points) overlaid on the $L_3$-edge powder XAS spectrum from Ref. \onlinecite{Hu2000} (light grey, filled curve). The red and blue curves in (b) represent partial energy-transfer integrations which are identified as inelastic signal derived from $t_{2g}$ and $e_g$ intermediate states by comparison to corresponding peaks in the Ru and O-site XAS spectra, indicated by vertical gray lines.  The axes in (a) and (b) are aligned on the $t_{2g}$-derived peaks at $529.6$ eV and $462.6$ eV, respectively.}
\label{fig:fig2}
\end{figure}

We report linear dichroic O $K$-edge XAS taken under incident $\sigma$ and $\pi$ polarizations in Fig. \ref{fig:fig2}(a). Besides the main edge onset ($E_i \simeq 537$ eV), two prominent pre-edge peaks arise from O-$2p$ -- Ru-$4d$ hybridization.  These represent the O-$2p$ projection of the unoccupied anti-bonding molecular orbitals based on $t_{2g}$ ($E_i \simeq 529.6$ eV) and $e_g$ ($E_i \simeq 532.2$ eV) Ru states. The separation between these peaks provides an estimated crystal field splitting between the \textit{unoccupied} $t_{2g}$ and $e_g$ states, which is $\Delta(\underline{e_g}-\underline{t_{2g}}) \approx 2.6$ eV as indicated in Fig. \ref{fig:fig2}.  We highlight the large degree of linear dichroism, particularly at the $t_{2g}$ hybridization pre-edge peak. Considering the global tetragonal lattice symmetry and the polarization projections in the crystal axes ($\sigma \parallel [0,1,0]$; $\pi \parallel [0.97, 0, 0.26]$), one finds a remarkable sensitivity to the in- vs. out-of-plane polarization component \cite{Das2018}. The origin of the large linear dichroism at the $t_{2g}$ resonance has been discussed in the context of other rutile oxides \cite{Kim2016,Stagarescu2000} and represents a partial quenching of the $\pi$-bonding strength for the $|yz\rangle$ and $|x^2-y^2\rangle$ orbitals, leaving the latter essentially non-bonding with respect to the O-$2p$ orbitals.  The origin of this effect is due to the longer-range connectivity of RuO$_6$ octahedra in rutile structure, where O-$2p$ orbitals that have the symmetry properties for $\pi$-bonding are instead activated in strong $\sigma$-bonds with neighboring Ru ions \cite{Sorantin1992}. This is in contrast to oxides of perovskite symmetry, where a unique separation of oxygen orbitals into global sets which are active in pure $\pi$- and $\sigma$-bonding is possible. These observations directly support a strong anisotropy in the Ru-O bonding properties, particularly with the $t_{2g}$ orbitals \cite{Goodenough1971a, Sorantin1992, Kahk2014,Das2018}.

%CNote that this is in contrast to typical perovskite-oxides where a unique global separation of O orbitals into sets active in pure $\pi$- and $\sigma$-bonding is possible. These observations directly support a strong anisotropy in the Ru-O bonding properties, particular with the $t_{2g}$ orbitals \cite{Goodenough1971a, Sorantin1992, Kahk2014,Das2018}.

In Fig. \ref{fig:fig2}(b), we report the Ru $M_3$-edge XAS spectrum collected with $\pi$ incident polarization.  The grey data points represent XAS signal in partial fluorescence yield (PFY) (energy transfer window $0.2 \to 10.0$ eV), which is plotted over the $L_3$-edge powder XAS spectrum (from Ref. \onlinecite{Hu2000}) shifted by the tabulated $L_3$-$M_3$ edge separation ($2376.6$ eV) \cite{Thompson2009}. We note both a strong agreement between the XAS profiles at the Ru $M$ and $L$ edges over the region measured, as well as a close correspondence of the characteristic two-peak structure at the Ru edges and O pre-edge.  The latter is highlighted by aligning the incident energy axes in Fig. \ref{fig:fig2}(a,b) to the lower energy $t_{2g}$-derived peaks (at energies 529.6 and 462.6 eV, respectively), indicated by the vertical grey lines. This corroborates the energy scale $\Delta(\underline{e_g}-\underline{t_{2g}}) \approx 2.6$ eV from both measurements. 

%Figure 3%

\begin{figure}[t]
\includegraphics[width=1.0\columnwidth]{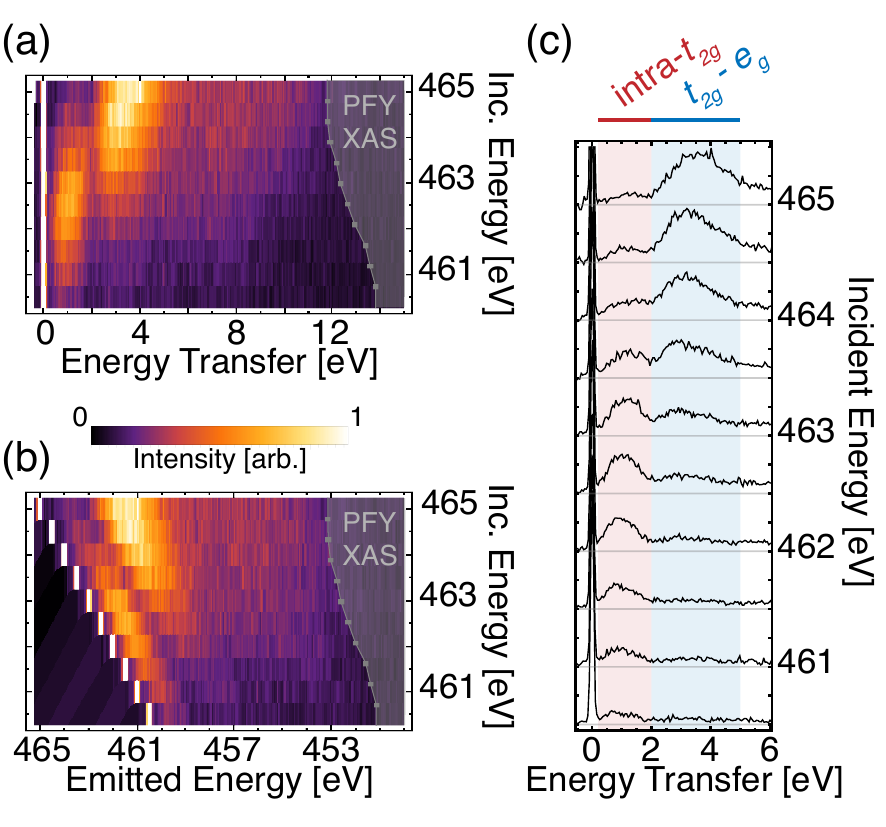}
\caption{Ru $M_3$-edge RIXS maps plotted against (a) energy transfer and (b) absolute emitted energy across the Ru $M_3$-edge.  Insets on the left axis in (a) denote the PFY XAS determined from integration of the inelastic signal as in Fig. \ref{fig:fig1}. (c) Individual line scans in the $dd$-excitation region for incident energy $E_i = 460.5$ eV (bottom) to $E_i = 465.0$ eV in steps of $0.5$ eV.  The regions corresponding to intra-$t_{2g}$ and $t_{2g}$-$e_g$ $dd$-excitations are demarcated by vertical gray lines.}
\label{fig:fig3}
\end{figure}

While XAS can assess the coarse crystal field energies (e.g. $10Dq$), they provide less direct insight into the $t_{2g}$ orbital energies of particular importance for RuO$_2$, a multi-orbital $t_{2g}$ system. To glean further information, we perform RIXS with incident $\pi$ polarization across the Ru $M_3$ edge (460.5-465 eV).  The resultant RIXS maps plotted against energy transfer and emitted photon energy are depicted in Fig. \ref{fig:fig3}(a) and (b), respectively. These measurements reveal two Raman-like $dd$-excitations around $1$ and $3.5$ eV which are respectively identified as intra-$t_{2g}$ and $t_{2g}$-$e_g$ excitations.  This identification is unequivocally confirmed by the incident-energy dependent intensity of these peaks, whose regions are defined in Fig. \ref{fig:fig3}(c) and plotted against the XAS in Fig. \ref{fig:fig2}(b).  We find a remarkably close correspondence, with the two excitations peaking with incidence energy at the $t_{2g}$ and $e_g$ intermediate states as suggested by the O $K$-edge and Ru $M_3$-edge XAS with a splitting of $2.6$ eV.  This splitting in incident energy is accompanied by a corresponding shift of the excitations along the energy transfer axis (Fig. \ref{fig:fig2}) by the same value.  These features directly confirm the origin of each peak, but raise an important question regarding the energy scale of the intra-$t_{2g}$ excitations ($\approx1$ eV) in the RIXS spectra. In $O_h$ symmetry, $t_{2g}$ levels are nearly degenerate and split solely through SOC (160 meV for Ru$^{4+}$) which cannot account for the observed splitting. This feature, along with the large Ru-O bonding anisotropy deduced from the O $K$-edge dichroism, is suggestive of an additional structural component breaking the $t_{2g}$ orbital degeneracies.

\section{Modeling}\label{modeling}

To test these predictions, one must recall the sensitivity of RIXS to the multi-electron multiplet structure in both the initial and intermediate RIXS state \cite{MorettiSala2011,He2016}.  Therefore, to resolve the interplay between low-symmetry crystal field and SOC, the effects of electron correlations must be properly included for a quantitative comparison to the RIXS spectra. To achieve this, we employ crystal field multiplet and core-level spectra calculations for the local $4d^4$ Ru$^{4+}$ ion as implemented in Quanty \cite{Haverkort2014,Haverkort2012,Lu2014,Lu2019,QuantyCite}. In the calculations, we include the full rutile crystal field, electronic correlations and SOC. Details for the calculations, including all parameters, are included in Appendix B. The calculations (Fig. \ref{fig:fig4}) account for the experimental geometry and polarization conditions.

%% FIGURE 4 %%

\begin{figure*}[t]
\includegraphics[width=\textwidth]{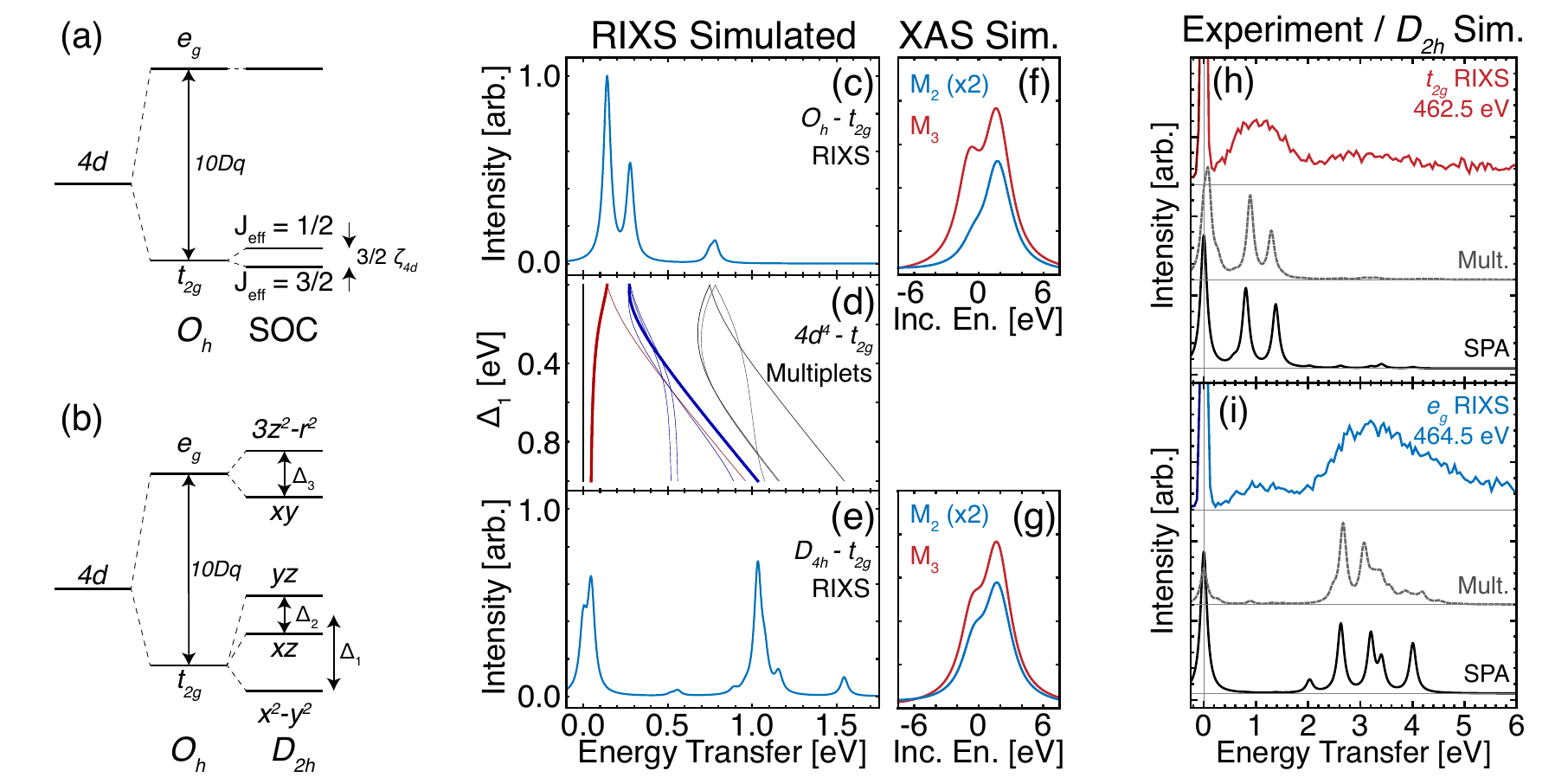}
\caption{Single particle energy level diagrams for the $4d$ orbitals are presented for (a) $O_h$ symmetry in the presence of SOC and (b) a reduced-symmetry $D_{2h}$ crystal field splitting.  In (b), the parameters $\Delta_{1-3}$ are defined to maintain the average configuration energy and the $4d$ orbitals are defined with respect to the local orthorhombic axes (see Fig. \ref{fig:fig1}(a,c)) for transition metal sites in the rutile structure  \cite{Rudowicz1992,Kahk2014,Ping2015,Mattheiss,Sorantin1992,Goodenough1971a}. The $t_{2g}$ RIXS spectrum for an $O_h$ model is shown in (c), depicting two dominant excitations around $200$ meV. These are identified as multiplet excitations with respect to the singlet ground state energy with the thick red and blue lines in (d), which shows the evolution of the $t_{2g}$ multiplet spectrum as $O_h$ symmetry is reduced to $D_{4h}$ with the CF parameter $\Delta_1:0 \to 1$ eV (see text for details). (e) The $t_{2g}$ RIXS spectrum for maximum $\Delta_1 = 1$ eV in the multiplet spectrum in (d).  The corresponding $M_{2,3}$ XAS curves for the $O_h$ and $D_{4h}$ models are presented in (f,g), respectively, with the $M_2$ spectrum intensity multiplied by 2 to highlight deviations from the statistical $M_{3}/M_{2}$ branching ratio.  Both edges are shifted in incident energy to the center of the configuration energy, defined as the energy zero. (h,i) Comparison between atomic multiplet (grey, dashed) and a corresponding single-particle crystal field model (black, solid) in $D_{2h}$ symmetry at the $t_{2g}$ ($E_i = 462.5$ eV) and $e_g$ ($E_i = 464.5$ eV) resonances, respectively.}
\label{fig:fig4}
\end{figure*}

%{\color{blue}the parameters $\Delta_{1-3}$ are defined to maintain the average configuration energy and the $4d$ orbitals are defined with respect to the local orthorhombic axes (see Fig. \ref{fig:fig1}(a),(c)) for transition metal sites in the rutile structure  \cite{Rudowicz1992,Kahk2014,Ping2015,Mattheiss,Sorantin1992,Goodenough1971a}.}

%In (b), the parameters $\Delta_{1-3}$ are defined to maintain to average configuration energy and orbital definitions respect the standard orthorhombic axes (see Fig. \ref{fig:fig1}(a),(c)) for transition metal sites in the rutile structure  \cite{Rudowicz1992,Kahk2014,Ping2015,Mattheiss,Sorantin1992,Goodenough1971a}.

\subsection{$O_h \to D_{4h}$ symmetry}

Beyond the cubic crystal field energy ($10Dq\sim2.6$ eV), it was argued that SOC in RuO$_2$ is the second dominant contribution with a negligible $D_{2h}$ crystal field component \cite{Hu2000,Hirata2013}.  In the single particle limit, this results in a SOC split $J_{\text{eff}} = \frac{3}{2}$, $\frac{1}{2}$ $t_{2g}$-subspace [see Fig. \ref{fig:fig4}(a)], which is a typical model for nearly octahedral $4d$ and $5d$ oxides \cite{Das2018,Gretarsson2019,Sala2014,Kim2008}. A consequence of this assumption is a quenching of the $|2p_{1/2}\rangle \to |t_{2g}\rangle$ channel in the $L_2$ XAS spectrum in the single particle limit. This necessitates a further inclusion of electron correlations to maintain agreement with the XAS doublet at the $L_{2}$ and $L_{3}$ edges \cite{Hu2000}. We note that the $J_\text{eff}$ scenario is successful in understanding the $M$/$L$-edge RIXS spectra of the more localized RuCl$_3$ \cite{Lebert2020, Suzuki2020}. 

We reformulate this model elaborated for the $L$-edges to the $M_3$ edge through an appropriate replacement of the intermediate state $pd$ correlation parameters.  This is required to describe the interaction of the valence electrons with the core-hole created after the absorption step in the RIXS process.  The $3p$ core-hole at the $M$-edges is shallower than the $2p$ core-hole of the $L$-edges, which modulates the interaction of this $p$ core-hole with the valence $4d$ electrons \cite{Miedema2019,Wray2015,Chiuzbaian2005}. The intra-atomic electron interactions amongst the valence electrons (`$dd$' correlations) and the intermediate state interaction of the valence $d$ electrons with the core $p$-shell (`$pd$' correlations), are introduced through the direct (Slater) and exchange Coulomb interaction integrals (see Appendix B).  These quantities are reduced from the atomic values (obtained through ab-initio Hartree-Fock calculations \cite{Haverkort2005a}) due to screening effects introduced in the solid state \cite{DeGroot2008,Haverkort2012,Hu2000}. For our model, we use a uniform screening of the $dd$ correlation parameters leading to a value of $40\%$ compared to atomic values.  For the $pd$ correlation parameters, which are screened less effectively \cite{DeGroot1994}, we use a decreased value of $60\%$ of atomic values. We report the expected RIXS spectrum within the intra-$t_{2g}$ excitation region in Fig. \ref{fig:fig4}(c). The dominant low-energy $dd$-excitations in the experimental polarization conditions are expected to arise near the $t_{2g}$:$|J = 3/2\rangle \to |J = 1/2\rangle$ with energy $\frac{3}{2}\zeta_{4d}$ ($\frac{3}{2}160$ meV$\approx$240 meV), rather than the much higher energy excitations observed ($\sim1$ eV).

%We reformulate this model to the $M_3$ edge in RuO$_2$ using a $40\%$ screening of the $dd$ Slater integrals from atomic values and a $60\%$ screening of the intermediate state $pd$ parameters, which are screened less effectively \cite{DeGroot1994}.%

%We have verified the results of Ref. \onlinecite{Hu2000}, where $dd$ correlations exceeding $40\%$ atomic values lose agreement with experiment quickly by introducing a strong mixing between the $t_{2g}$ and $e_g$ intermediate state resonances, destroying the characteristic double-peaked feature of the XAS spectrum at both edges.

We find that increasing the $dd$ Slater integral scaling above $40\%$ of atomic values leads to a loss of agreement with experiment, due to an overestimation of the multiplet effects for $4d$ oxides (see Appendix B and Ref. \onlinecite{Hu2000}). The primary effect is the loss of the characteristic double-peaked structure of the XAS spectrum at both the $M_2$/$M_3$ edges, due to a strong mixing between the $t_{2g}$ and $e_g$ intermediate state resonances. Furthermore, we find a low sensitivity of the XAS/RIXS spectra to the intermediate state $pd$ correlation scaling, which is due to the larger relative intermediate-state spin-orbit splitting between the $4d$ $M_{2,3}$ edges compared to $3d$ TM $L_{2,3}$ edges \cite{DeGroot1994}. This also implies only small quantitative differences between spectra at the $M_{2,3}$ and the $L_{2,3}$ edges.  We therefore fix the correlation parameters as typical values for Ru$^{4+}$ ions \cite{DeGroot1994} and further explore the $t_{2g}$ multiplet spectrum as the $O_h$ symmetry is reduced to $D_{4h}$ and report our results in Fig. \ref{fig:fig4}(d).  The multiplet energies are plotted with respect to the effective total angular momentum $J = 0$ singlet ground state as a function of increasing CF parameter $\Delta_1$ [see Fig. \ref{fig:fig4}(b)], raising the $d_{xz}/d_{yz}$ orbitals above the $d_{x^2-y^2}$ state.  The dominant excitations in the RIXS spectra are identified as transitions to two doublets with $S_z \simeq \pm 1$ (thick red line, $S_z $ - spin magnetic quantum number) and $L_z \simeq \pm 1$ (thick blue line, $L_z$ - orbital magnetic quantum number), which are split from a $J = 1$ triplet and $J = 2$ quintet in $O_h$ symmetry, respectively \cite{Gretarsson2019}. The remaining multiplets have substantially lower RIXS intensity. The RIXS spectrum for the endpoint ($\Delta_1 = 1$ eV) is plotted in Fig. \ref{fig:fig4}(e). We find that the dominant higher-energy intra-$t_{2g}$ multiplet excitation ($L_z$ doublet) is determined by the CF energy scale once the splitting exceeds the SOC coupling scale.  In contrast the low-energy, $S_z$ doublet excitation becomes very low in energy, $\sim 50$ meV for $\Delta_1 = 1$ eV.  
We conclude from these experimentally-accessible multiplet excitations that a large CF splitting in the $t_{2g}$ sector is required to reproduce the high intra-$t_{2g}$ excitation energy observed in RIXS experiment. With the introduction of this CF component, the $t_{2g}$ orbital energetics become dominantly determined by the CF as opposed to the SOC.  This is in contrast to the $J_\text{eff}$ model, where SOC is the only interaction breaking the $t_{2g}$ degeneracy.

%Furthermore, in this case, the energetics are determined dominantly by the CF splitting, in contrast to the SOC in the $J_\text{eff}$ model.
 
For the $O_h$ and $D_{4h}$ models in Fig. \ref{fig:fig4}(c,e), respectively, we also report the expected $M_{2,3}$ XAS spectra in Fig. \ref{fig:fig4}(f,g), respectively.  The $M_2$-edge in each case is multiplied by 2 in intensity to highlight deviations from the statistical $I(M_3)/I(M_2) = 2$ branching ratio (BR).  This $J_\text{eff}$ model leads to a very large BR $\sim3.75$, far in excess of the experimental value of 2.15 \cite{Hu2000}.  On the other hand, the $D_{4h}$ model maintains the double-peaked structure at both edges along with a substantial decrease in the BR to $\sim 2.75$, closer to experiment. Physically, the BR is related to the expectation value of the spin-orbit operator, $H_{SO} \propto \langle \mathbf{L} \cdot \mathbf{S} \rangle$, in the ground state \cite{Thole,Clancy2012}.  In $D_{4h}$ symmetry with $\Delta_1 = 1$ eV,  this expectation value is large for the $J = 0$ ground state ($H_{SO}\sim 1.19$ eV) while it is substantially lower in the low-lying $S_z$ doublet ($H_{SO} \sim 0.55$ eV).  Therefore, admixture of the $S_z$ doublet into the ground state by covalency/superexchange interactions may provide further reduction of the BR, while also endowing a finite magnetic moment at Ru sites. This is a known essential feature for understanding magnetism in Ru$^{4+}$ $t_{2g}$-systems with the formal $J = 0$ singlet (non-magnetic) ground state \cite{Khaliullin2013}.

%In this section, we have demonstrated that a large intra-$t_{2g}$ CF splitting is required to explain the large energy transfer of the intra-$t_{2g}$ RIXS excitation.  In particular, it is shown that the dominant RIXS features in experimental conditions are not derived from high-energy multiplet features in near octahedral symmetry, but rather lower-energy features related to SOC energetics in $O_h$ symmetry that are increased in energy by the $t_{2g}$ CF energy scale as the symmetry is lowered. Furthermore, this reduction in symmetry improves the agreement with the qualitative spectral features of the Ru $M$/$L$-edge XAS profile, and provides a more quantitatively accurate BR.  

\subsection{$D_{2h}$ symmetry}

With this information in hand, we turn to the optimization of the CF parameters in the full $D_{2h}$ symmetry with respect to the experimental results. We begin within the same multiplet model, with $\zeta_{4d} = 161$ meV and $dd$/$pd$ correlation scaling of $40\%$/$60\%$ atomic values, respectively.  We further employ the constraint that $10Dq = 2.6 + \Delta_1/3$, which ensures a splitting between the unoccupied $t_{2g}$ and $e_g$ states of $2.6$ eV in accordance with the Ru and ligand (O $K$) XAS (Fig. \ref{fig:fig2}). An optimal agreement is found with the parameters $\Delta_1 = 1.075$ eV, $\Delta_2 = 0.55$ eV and $\Delta_3 = 0.60$ eV.  The simulated RIXS spectra at the $t_{2g}$ and $e_g$ resonances in the multiplet model are compared to the experimental data in Fig. \ref{fig:fig4}(h,i) (dashed grey lines), respectively. We broadened the calculations by a Lorentzian of linewidth $0.1$eV to highlight the individual excitations.  We find a much better agreement at the intra-$t_{2g}$ excitation region compared to a near octahedral $J_\text{eff}$ model and a consistent behavior at the $t_{2g}$-$e_g$ excitation region.

In Fig. \ref{fig:fig4}(h,i), we also report an equivalent model of $dd$ CF transitions within the single particle approximation (SPA), shown as solid black lines.  These calculations are simulated by assuming the same CF parameters as the $D_{2h}$ symmetry multiplet model, with a corresponding single-particle energy diagram of Fig. \ref{fig:fig4}(b).  The SPA model is calculated with the intra-atomic $dd$ Slater integrals set to zero and a simulated spin-triplet ground state, with a $d^4$ filling of $(n_{x^2-y^2}, n_{xz}, n_{yz}) = (2, 1, 1)$ to agree with the $t_{2g}$ shell filling known from first-principles calculations \cite{Mattheiss,Berlijn2017}. The necessity to impose this ground state implies that RuO$_2$ is moderately correlated, due to the Hund's coupling that is required to stabilize the appropriate electron filling.  This assignment is in general agreement with transport and band-structure studies, which reveal only modest deviations due to electron-electron interactions \cite{Glassford1994, Ruf2020}. The use of the SPA calculations is employed here as a minimal model to capture the dominant CF interactions in the determination of the core-level spectra.  This eases the interpretation of the spectral features, as discussed below.

With the SPA, we find an improved consistency compared to the $D_{2h}$ multiplet model with respect to the high bandwidth of the $t_{2g}$-$e_g$ resonance and also a lack of the low-energy singlet $\to$ doublet transitions $\sim100$ meV which are not resolved within experiment. The SPA is a more natural description provided the distinct incident energy dependence, wherein higher energy $dd$ excitations resonate at higher incident energies (Fig. \ref{fig:fig3}) as well as for understanding the similar RIXS spectra available from metallic rutile-phase oxides, discussed in more details below.

%This model is calculated with quenched $dd$ correlations

To further examine the SPA interpretation in the context of RuO$_2$, we highlight the doublet structure at the intra-$t_{2g}$ RIXS resonance reported in Fig. \ref{fig:fig5}, where fits reveal a different resonant behavior of the two components (see Appendix A). We interpret these features as $d_{x^2-y^2} \to d_{xz}/d_{yz}$ $dd$ excitations, labelled as $(t_{2g})_{1,2}$ in Fig. \ref{fig:fig5}, respectively. The near-equal RIXS cross-section of these features is associated with the same (half-filled) occupancy of the $d_{xz}/d_{yz}$ orbitals. We particularly note that the splitting of the features in the energy transfer axis [Fig. \ref{fig:fig5}(a)] is nearly equal to the splitting of their resonances along the incidence energy axis [Fig. \ref{fig:fig5}(b)], a behavior consistent with the SPA model of $dd$-transitions between $D_{2h}$ CF levels [Fig. \ref{fig:fig5}(c)].  This incident energy dependence and a splitting on this energy scale is difficult to justify in the higher symmetry CF models [Fig. \ref{fig:fig4}]. These peak energies were used to fix the intra-$t_{2g}$ CF parameters in the $D_{2h}$ models above to $\Delta_1 = 1.075$ eV and $\Delta_2 = 0.55$ eV. The consideration of the experimental polarization condition confirms the $t_{2g}$ level ordering in Fig. \ref{fig:fig4}(b), in agreement with \textit{ab-initio} predictions \cite{Mattheiss}. In particular, we have $\Delta_1 > 0$, matching the expected destabilization of the $\pi$-bonding $t_{2g}$ orbitals with respect to the non-bonding $d_{x^2-y^2}$ state \cite{Sorantin1992}.

% We interpret these features as $d_{x^2-y^2} \to d_{xz}/d_{yz}$ transitions (labelled as $(t_{2g})_{1,2}$ in Fig. \ref{fig:fig5}, respectively) with a cross section due to equal occupancy of $d_{xz}/d_{yz}$ orbitals.

%% FIGURE 5 %%

\begin{figure}[t]
\includegraphics[width=\columnwidth]{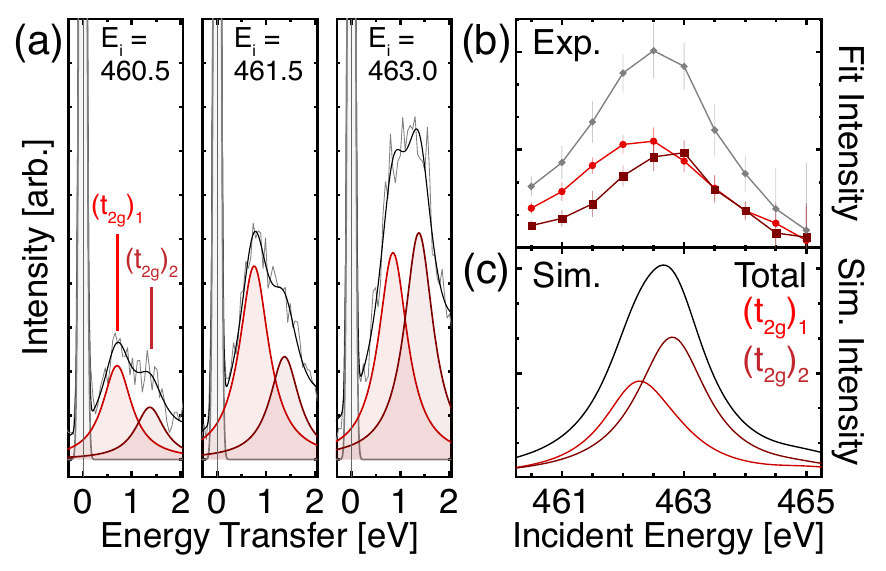}
\caption{(a) The intra-$t_{2g}$ region for selected incident energies including a fit (black line) to two Lorentzians (light/dark red) and a Gaussian elastic (grey shaded) over the experimental data (grey line).  The (b) fit amplitudes and (c) simulated amplitudes (from model of Fig. \ref{fig:fig3}(b)) as a function of incident energy, showing a clear splitting of the resonant energy position. Error bars denote the $80$\% confidence interval from the fits (see Appendix A).}
\label{fig:fig5}
\end{figure}

\subsection{Interpretation of crystal field parameters}

The local projection of the $D_{2h}$ CF components ($\Delta_{1-3}$) in RuO$_2$, as deduced from our RIXS measurements, are large compared to typical values in distorted octahedral environments commonly encountered in perovskite oxides.  In general, the CF splitting experienced in a solid can be decomposed into an ionic contribution, due to the symmetry of the local coordination of TM sites, and a covalent contribution due to the hybridization strength of the TM with the neighboring ligands \cite{Haverkort2012,Ushakov2011,Scaramucci2015}. We attribute these large CF contributions to the inherent difference in the bonding properties amongst the $t_{2g}$ orbitals, and therefore to a highly anisotropic covalent CF contribution. Evidence for this situation is provided by the large bonding anisotropy amongst the $t_{2g}$ orbitals as revealed by the large dichroism as the O $K$ pre-edge (Fig. \ref{fig:fig2}(a)). 

The appearance of a large CF splitting has been previously noted in the XAS spectrum of isostructural, rutile TiO$_2$ \cite{DeGroot1990} which was later attributed to a band structure effect appearing only for cluster sizes exceeding a full coordination of neighboring octahedra \cite{Kruger}.  The necessity for cluster sizes beyond the local MO$_6$ octahedron (M $=$ transition metal) to explain the full CF effects in rutile oxides can be equivalently viewed as introducing competition for bonding with shared O:$2p$ orbitals between neighboring TM sites. Therefore, we suggest that there is additional anisotropy in the covalent CF contribution beyond that suggested by the local environment alone which is the origin of the large intra-$t_{2g}$ energy between the non-bonding $d_{x^2-y^2}$ and $\pi$-bonding active $d_{xz}$/$d_{yz}$ orbitals, related to longer-range structural symmetry of the rutile lattice.  Finally, we note that through the interaction with bosonic excitations in the system (e.g. phonons \cite{Lee2014}), the $dd$-excitations can be broadened and shifted to apparently higher energy-transfer in RIXS spectra.  Therefore, the $4d$ orbital energies resolved in our RIXS experiments should be interpreted to represent an upper bound for the energies of the bare $4d$ CF levels.

\section{Discussion}\label{discussion}

\subsection{Implications for RuO$_2$}
From our evidence, we are able to fill in a gap concerning the electronic structure of RuO$_2$.  Overall, the spectra are characterized by a high-energy transfer intra-$t_{2g}$ resonance and a higher energy $t_{2g}$-$e_g$ excitation, indicating that SOC alone is insufficient for a quantitative interpretation of the RIXS data.  This directly supports a reduced-symmetry crystal field mechanism, as we have resolved in the model of Fig. \ref{fig:fig4}(h,i). However, one has to be careful as the dominance of the low symmetry splitting does not naturally imply the irrelevance of SOC. While our measurements reveal that the CF dominates the effects of SOC in the determination of the local $t_{2g}$ energy levels, the rutile structure hosts symmetry-protected band degeneracies (due to crystalline symmetries between sublattices) which are split \textit{only} by SOC \cite{Sun2017,Jovic2018}.  Therefore, SOC may have dramatic effects on transport properties (e.g. spin Hall conductivity) while playing only a modest role in the locally-probed orbital structure. The lack of identification of these states in RIXS may derive from their itinerant/delocalized nature which would produce incoherent fluorescence rather than sharp $dd$ excitations. 

%While SOC effects are effectively quenched by the crystal field at a local level,

Beyond SOC, the importance of the space and local point group symmetries has been discussed in several contexts, including the possible crystal Hall effect \cite{Smejkal2020}. The effect of this reduced symmetry in our $D_{2h}$ model shall be further used to refine the interpretation of the spectral lineshape in resonant diffraction experiments \cite{Zhu2019,Hirata2013} along with its connection to magnetic order.  Furthermore, the orbital energetics in connection to the lattice symmetry are of particular importance for interpreting the mechanism of strain-induced superconductivity in thin films of RuO$_2$, whether due to modulation of the electronic structure \cite{Ruf2020}, lattice instability/enhanced electron-phonon coupling \cite{Uchida2020}, or both. The local $4d$-orbital level energies have a direct connection to particular Ru-O and Ru-Ru bond strengths within a molecular bonding interpretation \cite{Goodenough1971,Sorantin1992} making the system prone to modification of lattice parameters through strain. 
Our work directly supports this molecular orbital picture. The Ru-O hybridization yields significant bonding anisotropy and breaks orbital degeneracies, the effects of which are directly probed by our measurements through their projection onto both the O-$2p$ and Ru-$4d$ states as measured through O-$K$ XAS and Ru-$M_3$ XAS/RIXS, respectively. Therefore, the experiments presented here are needed to assess the modification of the orbital energies under external perturbation such as strain.

%Our work directly supports this interpretation, through the Ru-O hybridization (O $K$-edge XAS) and the projection of these hybridization effects directly onto the Ru states (Ru $M_3$-edge RIXS/XAS).

\subsection{Comparison among metallic states in rutile oxides}

\begin{figure}[t]
\includegraphics[width=\columnwidth]{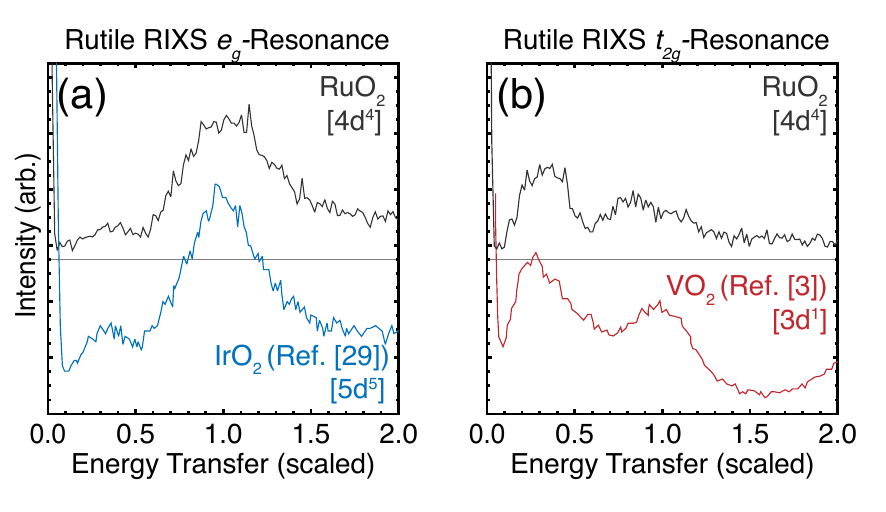}
\caption{RIXS spectra from metallic rutile oxides are compared.  (a) Comparison between Ir $L_3$-edge RIXS in IrO$_2$ [$5d^5$] (from Ref. \onlinecite{Kim2018}) and RuO$_2$ RIXS ($E_i = 465.0$ eV) at the $e_g$ final state resonance.  (b) Comparison between V $L_3$-edge RIXS in  VO$_2$ [$3d^1$] (from Ref. \onlinecite{He2016}) and RuO$_2$ RIXS ($E_i = 462.5$ eV) at the $t_{2g}$ final state resonance.  The energy transfer axis in each case is scaled by the approximate $t_{2g}$-$e_g$ excitation energy: $3.9$/$3.5$/$2.4$ eV for Ir/Ru/V, respectively.}
\label{fig:fig6}
\end{figure}

It is worth mentioning the similarity among the RIXS spectra in the metallic phase of the rutile oxides VO$_2$, RuO$_2$, and IrO$_2$ \cite{He2016,Kim2018}.  The available RIXS data for VO$_2$ \cite{He2016} and IrO$_2$ \cite{Kim2018} are near the $t_{2g}$/$e_{g}$ final-state resonances of the V/Ir $L_3$ edges, respectively.  These are compared to the Ru $M_3$-edge scans at the corresponding resonances ($t_{2g}$: $E_i = 462.5$ eV, $e_g$: $E_i = 465.0$ eV) in Fig. \ref{fig:fig6}.  The energy-transfer axes for the RIXS spectra in Fig. \ref{fig:fig6} are normalized by the respective energies of the high-energy $t_{2g} \to e_g$ $dd$-excitation. The increase of the $dd$-excitation energies across the V/Ru/Ir series is attributed to larger radial extension of higher $nd$ orbitals and the resulting increase in CF energy scales \cite{Khomskii2020}. In all cases, we find the spectra are characterized by a high-energy transfer intra-$t_{2g}$ excitation and a higher-energy $t_{2g}$-$e_g$ excitation, all with similarly broad $dd$-excitation linewidth. This is irrespective of the distinct $3d$/$4d$/$5d$ nature of each system and, therefore, the broad range of SOC strength [$30$meV (V), $160$meV (Ru), $500$meV (Ir)]. Given the smooth energetic trend of the intra-$t_{2g}$ and the $t_{2g}$-$e_g$ features across all systems, SOC alone is insufficient for a quantitative interpretation of the data.  This directly supports a common reduced-symmetry CF mechanism in all cases, as we have unambiguously resolved in the model of Fig. \ref{fig:fig4}(h,i), consistent with the interpretation in VO$_2$ \cite{He2016}. 

%The spectra in Fig. \ref{fig:fig6} are normalized along the energy-transfer axis to the center of the $t_{2g}$-$e_g$ excitation, with the increase of the $dd$-excitation energies across the V/Ru/Ir series attributed to larger radial extension of higher $nd$ orbitals and the commensurate increase in CF energy scales \cite{Khomskii2020}.

The consistency of the spectra also spans a wide range of formal $d$-electron count ($d^{1}-d^4-d^5$ for V-Ru-Ir), which have dramatically distinct multiplet structures that are not clearly reflected in experiments.  This strengthens the SPA interpretation of $dd$-excitations between CF levels introduced in the modelling above. The single-particle behavior may derive from the itinerancy of the higher-energy $d$-orbitals - except the non-bonding $d_{x^2-y^2}$ level - in rutile systems. This characteristic mixture of localized/itinerant electronic states may lead to a spectrum of partial excitations, where the emission stage of the coherent RIXS process is only active from the $d_{x^2-y^2}$ state, which forms a sharp peak in the occupied density of states \cite{Berlijn2017}.  In this case, other pathways would be expected to be dominated by incoherent decay channels between broad, itinerant bands resulting in an unresolvable continuum of particle-hole excitations \cite{Monney2012,Monney2020}. Such a contribution is potentially evidenced by the reduction of quasi-elastic spectral weight in the RIXS spectrum of VO$_2$ when crossing the MIT \cite{He2016} which is concomitant with an enhanced $d_{x^2-y^2}$ orbital polarization \cite{Haverkort2005}. This scenario would explain well the apparent lack of intra-atomic correlation effects in rutile RIXS spectra as in our supported model of Fig. \ref{fig:fig4}(h,i). Importantly, this model traces the characteristic RIXS response in rutile oxides to the inherent orbital anisotropy of the rutile structure and the the differential degree of covalent CF splitting among the $t_{2g}$ orbital states, providing additional support for our interpretation here for RuO$_2$.

\section{Conclusion}\label{conclusion}
In conclusion, we have measured Ru $M_3$ edge RIXS and O $K$-edge XAS linear dichroism in RuO$_2$.  Through the detection of the $dd$-excitation spectrum and multiplet modelling, our results firmly establish the dominance of low-symmetry crystal field in the local electronic structure.  This hierarchy is tightly bound to the rutile structure and its octahedral distortions and connectivity which indicates the need for a different treatment than conventional perovskite-based Ru compounds (such as Sr$_{n+1}$Ru$_n$O$_{3n+1}$). In RuO$_2$, the intermediate nature of the $4d^4$ Ru configuration highlights characteristic features of the RIXS spectra in rutile oxides, corroborating the universal role of the unique and often overlooked crystal field levels across the rutile family.  This represents a key discovery in the interplay of SOC and low-symmetry CF in $4d$/$5d$ oxides and is an essential step toward resolving the mechanisms of the novel physical properties (including the recently discovered superconductivity) in RuO$_2$.

\begin{acknowledgments}
We gratefully acknowledge Robert Green and Frank de Groot for insightful discussions. Work at MIT was supported by the Air Force Office of Scientific Research Young Investigator Program under grant FA9550-19-1-0063. Work at Brookhaven National Laboratory was supported by the DOE Office of Science under Contract No. DE-SC0012704. This work was supported by the U.S. Department of Energy (DOE) Office of Science, Early Career Research Program. This work was supported by the Laboratory Directed Research and Development project of Brookhaven National Laboratory No. 21-037.  The work at Argonne National Laboratory (crystal synthesis and pre-characterizations) was supported by the U.S. Department of Energy, Office of Science, Basic Energy Sciences, Materials Science and Engineering Division. This research used beamline 2-ID of the National Synchrotron Light Source II, a U.S. Department of Energy (DOE) Office of Science User Facility operated for the DOE Office of Science by Brookhaven National Laboratory under Contract No. DE-SC0012704.
\end{acknowledgments}

\appendix

\section{Fit Results for Experimental Data}
The incident energy dependent RIXS spectra presented in the main text were fit to a model consisting of a Gaussian elastic line, 2 Lorentzian curves in the intra-$t_{2g}$ region, 4 Lorentzian curves in the $t_{2g}-e_g$ multi-peaked energy region and a broader Lorentzian feature above $\Delta E = 5$eV energy transfer. The identification of two excitations at the intra-$t_{2g}$ region is motivated by (i) the clear transfer of spectral weight to higher energy transfer at higher incident energies above the $t_{2g}$ resonance [Fig. \ref{fig:fig2} c)], and (ii) the clearly resolved double peak structure for incident energies below the $t_{2g}$ final state resonance [$E_i = 462.5$ eV, see in particular Fig. \ref{fig:fig5} (b)]. Due to the large core-hole broadening ($\sim 1$-$2$ eV) along the incident energy axis, the double peaked structure must persist over a broad range of incident energies.  Therefore, the same model was employed to fit all incident energies with the same initial values and parameter restrictions, with freedom in the energy transfer position and amplitude.  The linewidth of the intra-$t_{2g}$ Lorentzians were restricted to be equal, with a similar restriction for the $t_{2g}-e_g$ peaks.  

\begin{figure}[t]
\includegraphics[width=\columnwidth]{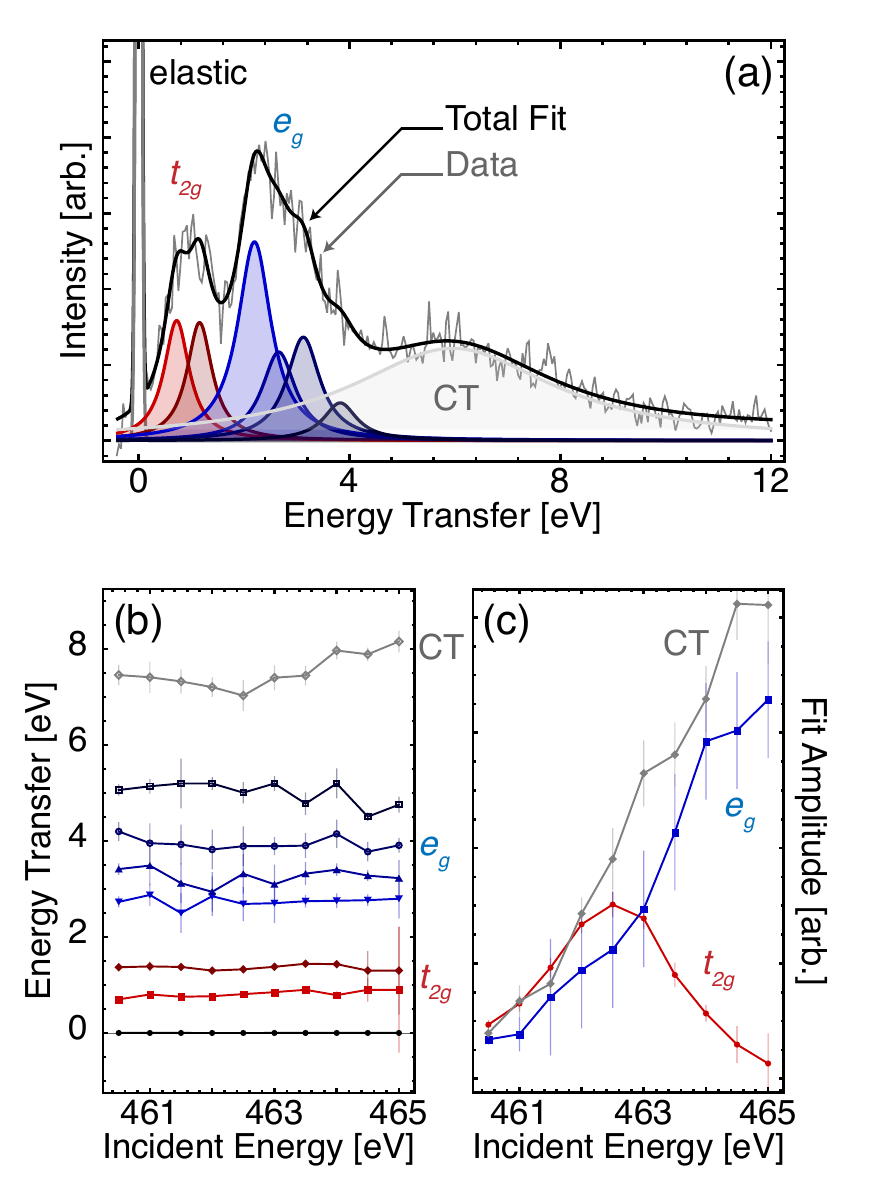}
\caption{(a) Representative data (gray) at $E_i = 463.5$eV compared to the total fit (black) and individual fit contributions as described in the text.  Contributions include the elastic (dark gray), intra-$t_{2g}$ $dd$ excitations (red), $t_{2g}$-$e_g$ excitations (blue) and the charge transfer (light gray).  (b) Incident energy dependent peak positions as fit from contributions delineated in (a).  The sum intensity of all $t_{2g}$, $e_g$ and charge transfer excitation peaks are compared in (c).  All error bars delineate 80\% confidence intervals in the fit parameters.}
\label{fig:figS1}
\end{figure}

The supported crystal field model with reduced symmetry described in the main text [Fig. \ref{fig:fig4}(h,i)] predicts additional peaks in the $t_{2g}-e_g$ region (2.0 to 5.2 eV energy transfer), which are not immediately resolvable in the data due to linewidths significantly in excess of the measurement resolution and overlap in energy transfer.  Despite not being able to resolve all spectral features, we attempt to fit this energy loss region using the minimal number of curves and present the resulting energy positions in Fig. \ref{fig:figS1}(b).  Such energy positions in this energy window must be interpreted keeping in mind the limitations of the data, set by the intrinsically broad linewidth of the $dd$-excitations. A representative fit for $E_i = 463.5$eV is shown in Fig. \ref{fig:figS1}(a). The fit full width at half maximum (FWHM) of the elastic line is $\approx\Delta E \simeq 125$meV, and nearly constant as a function of incident energy.  We therefore conclude that the elastic line in the experimental data is completely dominated by the diffuse scattering of the incident beam and we find no evidence of quasielastic signal in the measurement.

In Fig. \ref{fig:figS1}(b), we summarize the peak positions in energy transfer versus incident energy for the individual contributions highlighted in Fig. \ref{fig:figS1}(a).  The intra-$t_{2g}$ peaks are discussed in the main text.   At the pre-edge region ($E_i < 462.5$eV), the fit suggests a Raman-like peak around $\Delta E = 7$eV, which begins to crossover to fluorescence-like behavior above $E_i = 462.5$eV.  In Fig. \ref{fig:figS1}(c), we display the intensity of the total intra-$t_{2g}$, $t_{2g}-e_g$, and charge transfer fits as a function of incident energy.  As can be seen, the intensity of the charge transfer transition tracks closely with the increase of the $t_{2g}-e_g$ intensity.  Furthermore, the crossover point from Raman- to fluorescence-like behavior is concomitant with the resonance position of the intra-$t_{2g}$ $dd$-excitation.  The incident energy dependence from the fits for the $t_{2g}$ and $e_g$ excitations agree well with the energy-transfer window integrations used for PFY analysis presented in Fig. \ref{fig:fig2} in the main text.

\section{Crystal Field Multiplet Calculations}

Core-level X-Ray Absorption and Resonant Inelastic X-Ray Scattering spectra at the Ru $M_3$-edge are calculated with crystal field multiplet calculations as implemented in the Quanty software \cite{Haverkort2014,Haverkort2012,Lu2014,Lu2019,QuantyCite}.  The main idea of this approach is to model the system as atomic Ru$^{4+}$, with modifications to the Hamiltonian to account for the effects of the crystalline environment \cite{DeGroot2008}. Such an approach permits an exact treatment of both multielectronic effects and the fully relativistic core-levels, both of which are essential for the description the $L_{2,3}$/$M_{2,3}$ edges of transition metal systems \cite{DeGroot1994,Haverkort2012}. For the calculations considered here, we use a basis consisting of the Ru-$4d$ shell (10 single-particle states: 5 $d$-orbitals + spin up/down), occupied by the nominal four valence electrons.  For calculations of spectra at the $M_3$ edge, involving a dipole transition between the Ru $3p$ core-level and the Ru-$4d$ level, we also include the Ru $3p$ shell into the basis.  The oxygen $2p$ orbitals are not explicitly accounted for in the calculations.  Instead, the crystalline environment is modeled through a single-electron crystal field contribution to the Hamiltonian which obeys the local site-symmetry (orthorhombic, $D_{2h}$) of the Ru ions induced by the coordination of oxygen ligands in the rutile structure.

The multielectron ground state ($3p^6\;4d^4$) is determined by exact diagonalization within the Ru $4d$ shell, considering the SOC ($\zeta_{4d}$), the crystal field and the intra-atomic electronic correlations amongst the $4d$ electrons.  The latter are parametrized through the Slater integrals: $F^0(dd)$, $F^2(dd)$, and $F^4(dd)$.  In the RIXS intermediate state ($3p^5\;4d^5$), there is an additional contribution that considers the Coulomb interaction between the $3p$ core-hole and the valence $4d$ electrons which are parametrized by the Slater and exchange integrals $F^2(dd)$, $G^1(dd)$ and $G^3(dd)$.  These multielectron interactions are used as defined in the theory of atomic spectra \cite{Cowan1981}. For the calculation of both XAS and RIXS spectra, we consider dipole ($3p \to 4d$) transitions using dipole operators that reflect the experimental polarization conditions.

The crystalline environment of the Ru ions is modeled as a phenomenological average introduced through the crystal field, as well as a reduction (or screening) of both the $dd$ and $pd$ intra-atomic correlation parameters from atomic values.  Such a screening of the electronic correlations from atomic values is a well-known procedure for calculating the core-level spectra of transition metal ions in crystalline environments \cite{DeGroot1994,DeGroot2008,Hu2000,Haverkort2012}.  In our case, we use screened values of the $dd$/$pd$ Slater and exchange integrals to $40\%/60\%$ of atomic values, respectively.  These values for the screened parameters represent an established parameter regime for $4d$ transition metal oxides \cite{Hu2000,DeGroot1994,Gretarsson2019}.  For the SOC, the atomic value for the Ru$^{4+}$ ion ($\zeta_{4d} = 161$ meV) is used. We provide below in Table \ref{tab:tabS1} the values of all parameters used for calculations at the $M_{2,3}$ edges, taken from the thesis of M. Haverkort \cite{Haverkort2005a} and the Crispy database \cite{retegan_crispy}.  These values represent the bare Hartree-Fock values.

\begin{table}[h]
\caption{\label{tab:tabS1} CFM Model Parameters (values in eV)}
\begin{ruledtabular}
\begin{tabular}{c|c|c|c|c}
Init. State & $\zeta_{4d}$ & $F^{(2)}(dd)$ & $F^{(4)}(dd)$ & - \\
\hline
Ru $4d^4$ & 0.161 & 9.211 & 6.093 & -\\
\hline\hline
Inter. State & $\zeta_{3p}$ & $F^{(2)}(pd)$ & $G^{(1)}(pd)$ & $G^{(3)}(pd)$ \\
\hline 
Ru $3p^54d^5$ &    14.999   &    4.971      &     1.060     &      1.071      \\
\end{tabular}
\end{ruledtabular}
\end{table}

The crystal field acting on the Ru 4d shell is introduced through a single-electron contribution to the Hamiltonian in both the ground and intermediate states which defines the single-particle energies for the different real 4d orbitals.  Since the oxygen ligands are not explicitly accounted for in our calculations, the crystal field values assigned represent the total of the ionic and covalent contributions to the splittings \cite{Haverkort2014}, which in general are separable based on the bare Coulomb contribution and the contribution due to hybridization \cite{Ushakov2011,Scaramucci2015}, as discussed in the main text. We stress this fact so the parameters may be interpreted accordingly. For modelling the crystal field, we define a distinct set of parameters from typical approaches, which is more well-suited to the splittings observed in the rutile symmetry.  First, $'10Dq'$ is defined so that the $e_g$ states ($d_{3z^2-r^2}$ and $d_{xy}$) are raised in energy $\frac{3}{5} 10Dq$, while the $t_{2g}$ levels ($d_{x^2-y^2}$, $d_{xz}$, $d_{yz}$) are lowered by $-\frac{2}{5} 10Dq$.  As a note, the $d_{x^2-y^2}$ and $d_{xy}$ are switched with respect to typical definitions of the $e_g$ and $t_{2g}$ orbitals.  This is due to the principle axes of the orthorhombic crystal field being oriented $45^\circ$ away from the standard axes [see Main Fig. \ref{fig:fig1}(a)].  This is consistent to prior discussions of the orbital projections for rutile oxides, albeit sometimes with different notations \cite{Mattheiss,Kahk2014,Ping2015,He2016}.  Further crystal field parameters are defined by $\Delta_{1-3}$ for deviations from the octahedral symmetry.  $\Delta_1$ is the splitting between non-bonding $d_{x^2-y^2}$ and the $\pi$-bonding $d_{xz}/d_{yz}$ orbitals, while $\Delta_2$ is the splitting between the latter subspace. $\Delta_3$ is defined as the splitting between the $e_g$ orbitals.  

Overall the onsite energies are as thus, defined so that the average $4d$ configuration energy is constant with respect to each parameter:

\begin{align*}
E_{3z^2-r^2} &= \phantom{-}\frac{3}{5} 10Dq + \frac{\Delta_3}{2}\\
E_{xz} &= \phantom{-}\frac{3}{5} 10Dq - \frac{\Delta_3}{2}\\
E_{yz} &= -\frac{2}{5} 10Dq + \frac{\Delta_1}{3} + \frac{\Delta_2}{2}\\
E_{xz} &= -\frac{2}{5} 10Dq + \frac{\Delta_1}{3} - \frac{\Delta_2}{2}\\
E_{x^2-y^2} &= -\frac{2}{5} 10Dq - \frac{2}{3} \Delta_1 
\end{align*}

The optimized parameters for the $D_{2h}$ crystal field model in Fig. 3(b) of the main text are $\Delta_1 = 1.075$ eV and $\Delta_2 = 0.55$ eV, restricted by the energy of the intra-$t_{2g}$ doublet in the RIXS spectrum (main text Fig. \ref{fig:fig5} and Fig. \ref{fig:figS1}), $\Delta_3 = 0.6$eV and $10Dq = 2.6 + \frac{\Delta_1}{3} = 2.96$eV, the latter being restricted by the unoccupied state splitting as suggested by the resonant energy dependence and the Ru $L$/$M$-edge/O $K$-edge absorption profile.  Our sensitivity to $\Delta_3$ is less restrictive compared to other parameters, due to the many overlapping, broad peaks at the $t_{2g}-e_g$ excitations.  This parameter was optimized to best reproduce the broad bandwidth of the $t_{2g}-e_g$ features without creating additional, lower-energy excitations of a $d_{xz}/d_{yz} \to e_g$ nature that occur below the noticeable gap between the intra-$t_{2g}$ and $t_{2g}-e_g$ features around $\Delta E = 2$ eV.

%\bibliography{RuO2_Main_20201229.bib}
\end{document}